\documentclass{article}                                               
\usepackage{amssymb}
\renewcommand{\d}{\mathrm{d}}

\title
{Quantization of the $N=2$ Supersymmetric KdV Hierarchy}
\author{Anton M. Zeitlin\\
St. Petersburg Department of Steklov Mathematical 
Institute,\\ 
Fontanka, 27, St. Petersburg, 191023, Russia\\
zam@math.ipme.ru, http://www.ipme.ru/zam.html}
\begin {document}

\maketitle
\begin{abstract}
We continue the study of the quantization of 
supersymmetric integrable KdV hierarchies. We consider 
the N=2 KdV model based on the $sl^{(1)}(2|1)$ affine algebra but with 
a new algebraic construction for the L-operator, different from
the standard Drinfeld-Sokolov reduction. 
We construct the quantum monodromy matrix  
satisfying a special version of the reflection equation and show 
that in the classical limit, this object gives 
the monodromy matrix of N=2 supersymmetric KdV system. 
We also show that at both the classical and the quantum levels, 
the trace of the monodromy matrix (transfer matrix) 
is invariant under two supersymmetry transformations and the zero mode  of the associated U(1) 
current.       
\end{abstract}

{\bf Keywords:} superconformal field theory, quantum superalgebras, supersymmetric KdV equation,
supersymmetric integrable systems, quantization.

\section{Introduction}
In \cite{super-kdv}, \cite{toda-kdv2} we considered the quantization of the 
Drinfeld-Sokolov hierarchies associated with affine superalgebras along the lines of the bosonic 
approach introduced in \cite{blz} for the usual $\hat{sl}(2)$ KdV model. 
Here, we extend this approach to nonstandard KdV hierarchies related to the $sl^{(1)}(m+1|m)$ 
superalgebras. We consider the N=2 KdV model with the underlying affine superalgebra 
$sl^{(1)}(2|1)$ in detail, but the generalization to the higher rank case is straightforward. 
The interest in such 
integrable models arises because the quantization 
procedure becomes more involved than for the standard Drinfeld-Sokolov hierarchies.

To begin quantizing this model, we first consider  the corresponding classical theory 
(see Sec. 2). 
The classical version of the associated monodromy matrix is represented using the familiar 
P-exponential, but the exponent consists not only of generators corresponding to simple roots and 
their quadratic combinations but also of  generators corresponding to more complicated 
composite roots (see Sec. 2). 
We prove that all these terms corresponding to composite roots
(as in the ``standard'' case)
disappear from the first iteration of the quantum generalization of the P-exponential.
In addition, we show that 
this ``quantum'' P-exponential coincides with the reduced universal R-matrix of $\hat{sl}_q(2|1)$ quantum affine 
superalgebra (see Sec. 3 and 4).
We also prove that  supertraces of the quantum version of the monodromy matrix,
the so-called transfer matrices (by an obvious analogy with the lattice case), commute with the 
supersymmetry generators. Hence, these generators can be included in the family of  integrals of motion 
at both the classical and the quantum levels (see Sec. 5). 
In the last section, we also discuss  the relation of N=2  supersymmetric KdV model to the topological theories 
and their integrable perturbations.

\section{The $N=2$ SUSY KdV hierarchy in nonstandard form}
The matrix L-operator of the N=2 SUSY KdV hierarchy has the  explicit form \cite{delduc}:
\begin{eqnarray}
%\begin{displaymath}
\mathcal{L}_F=D-
\left(\begin{array}{ccc}
D\Phi_1 & 1 & 0  \\
\lambda & D\Phi_1-D\Phi_2 & 1\\
\lambda D(\Phi_1- \Phi_2)& \lambda & -D\Phi_2\\
\end{array}\right),
%\end{displaymath}
\end{eqnarray}
where $D=\partial_{\theta}+\theta\partial_u$, $u$ is a variable on a cylinder of circumference 
$2\pi$,
$\theta$ is a Grassmann number,
 and $\Phi_i(u,\theta)=\phi_i(u)-\frac{i}{\sqrt{2}}\theta\xi_i(u)$ are the superfields with the  
Poisson brackets
\begin{eqnarray}
&&\{\Phi_i(u_1,\theta_1),\Phi_i(u_2,\theta_2)\}=0\quad (i=1,2)\\
&&\{D_{u_1,\theta_1}\Phi_1(u_1,\theta_1),D_{u_2,\theta_2}
\Phi_2(u_2,\theta_2)\}=-D_{u_1,\theta_1}(\delta(u_1-u_2)(\theta_1-\theta_2))\nonumber
\end{eqnarray} 
This is the odd L-operator related to the $sl^{(1)}(2|1)$ affine superalgebra 
(the elements in the representation column vector
are graded top down in the order  ``even, odd, even''). 
There exists another (canonical) operator, also corresponding  to the N=2 SUSY KdV model,
related to the $sl^{(1)}(2|2)$ superalgebra \cite{inami}. 
This realization, which corresponds to 
a higher rank, can be quantized using  the procedure outlined in \cite{toda-kdv}, \cite{toda-kdv2},
which adds nothing new to the quantization procedure compared with \cite{super-kdv}-\cite{toda-kdv2}. 
Moreover, the representation theory of  $sl^{(1)}(2|2)$ is quite complicated at both
the classical and the quantum levels.

The form of the L-operator given above, corresponding to a  lower rank, allows working  with
the much  simpler representation theory 
of the $sl^{(1)}(2|1)$ superalgebra and provides very interesting features of the 
quantization formalism.

To build the scalar L-operator related to the matrix operator, 
we consider the 
linear problem $\mathcal{L}\Psi=0$ and express the second and third elements in the 
vector $\Psi$ in terms of the first (top) one. 
The linear equation for this element is
\begin{eqnarray} 
((D+D\Phi_1)(D-D(\Phi_1-\Phi_2))(D-D\Phi_2)+2\lambda D)\Psi_1=0.
\end{eqnarray}
We thus obtain  the scalar linear L-operator as
\begin{eqnarray}\label{fermil} 
L=D^3+(\mathcal{V}+2\lambda)D+\mathcal{U},
\end{eqnarray}
where the Miura map is specified by the relations
\begin{eqnarray}\label{pot} 
&&\mathcal{V}=-D\Phi_2 D\Phi_1+\partial\Phi_2-\partial\Phi_1=
\frac{i}{\sqrt{2}}\theta(\alpha^+-\alpha^-)  +V,\nonumber\\
&&\mathcal{U}=-\partial\Phi_2 D\Phi_1-D\partial\Phi_1=\theta U+\frac{i}{\sqrt{2}}\alpha^+,\\
&&U=-\phi_1'\phi_2'-\frac{1}{2}\xi_1\xi_2'-\phi_1'', \quad \alpha^+=\xi_1\phi_2'+\xi_1',\nonumber\\ 
&&\alpha^-=\xi_2\phi_1'+\xi_2', \quad V=\phi_2'-\phi_1'+\frac{1}{2}\xi_2\xi_1 \nonumber
\end{eqnarray}  
and $U,V$, and $\alpha^{\pm}$ satisfy the N=2 superconformal algebra under the 
Poisson brackets:
\begin{eqnarray}\label{poisson} 
&&\{U(u),\alpha^+(v)\}=-\alpha'^+(u)\delta(u-v)-2\alpha^+(u)\delta'(u-v),\nonumber\\
&&\{U(u),\alpha^-(v)\}=-\alpha^-(u)\delta'(u-v),\nonumber\\
&&\{\alpha^+(u),\alpha^-(v)\}=-2U(u)\delta(u-v)-2V(u)\delta'(u-v)-2\delta''(u-v),\nonumber\\
&&\{V(u), \alpha^+(v)\}=-\alpha^+(u) \delta(u-v),\nonumber\\
&&\{V(u), \alpha^-(v)\}=\alpha^-(u) \delta(u-v),\nonumber\\
&&\{U(u),V(v)\}=-V(u)\delta'(u-v),\nonumber\\
&&\{V(u),V(v)\}=-2\delta'(u-v),\nonumber\\
&&\{U(u),U(v)\}=-U'(u)\delta(u-v)-2U(u)\delta'(u-v).
\end{eqnarray}  
We now rewrite this L-operator in the algebraic form as 
\begin{eqnarray}
&&\mathcal{L}_F=D-(h_{\alpha_1}D\Phi_1+h_{\alpha_2}D\Phi_2+e_{\alpha_1}+e_{\alpha_2}+[e_{\alpha_2},
e_{\alpha_0}]+ \nonumber\\
&&[e_{\alpha_0}, e_{\alpha_1}]+D(\Phi_1- \Phi_2)e_{\alpha_0})
\end{eqnarray}
where $h_{\alpha_i}$ and $e_{\alpha_i}$ are the generators of the upper Borel algebra of  
$sl^{(1)}(2|1)$ with the commutation relations  
\begin{eqnarray}
&&[e_{\alpha_i}, e_{-\alpha_j}]=\delta_{i,j}h_{\alpha_i}\quad (i=0,1,2), \quad
[h_{\alpha_j},e_{\pm\alpha_i}]=\pm e_{\pm\alpha_i} \quad (i=1,2, i \neq j)\nonumber\\
&&[h_{\alpha_j},e_{\pm\alpha_0}]=\mp e_{\pm\alpha_0} \quad (i=1,2),\quad
[h_{\alpha_0},e_{\pm\alpha_i}]=\mp e_{\pm\alpha_i} \quad (i=1,2)\nonumber\\
&&[h_{\alpha_i},e_{\pm\alpha_i}]=0 \quad (i=1,2),\quad
[h_{\alpha_0},e_{\pm\alpha_0}]=\pm 2e_{\pm\alpha_0},\nonumber\\
&&ad^2_{e_{\pm\alpha_i}}e_{\pm\alpha_j}=0,\quad [e_{\pm\alpha_k},e_{\pm\alpha_k}] =0 \quad (k=1,2)
\end{eqnarray}
where the generators $e_{\pm\alpha_i}$ are odd for $i=1,2$ and even for $i=0$ and $[,]$ denotes the 
supercommutator.

The symmetrized Cartan matrix $b_{ij}=(\alpha_i,\alpha_j)$ corresponding 
to the given affine superalgebra is
\begin{eqnarray}
b=
\left(\begin{array}{ccc}
0 & 1 & -1  \\
1 & 0 & -1\\
-1& -1 & 2\\
\end{array}\right).
\end{eqnarray}
In this paper, we use only the evaluation representations 
(for which the central charge of the corresponding affine algebra is equal to zero).
 
To write a monodromy matrix, we must consider the equivalent bosonic L-operator. 
Expressing the linear problem associated with operator (\ref{fermil}) as
$$ 
\mathcal{L}\Psi=(\partial_{\theta}+\theta\partial_u+N_1+\theta N_0)(\Psi_0+\theta\Psi_1),
$$ 
we can easily rewrite it as a bosonic linear problem for $\Psi_0$: 
$$
\mathcal{L}_B\Psi_0\equiv
(\partial_u+N^2_1+N_0)\Psi_0=0
$$ 
with  $\Psi_1=-N_1 \Psi_0$,
where 
\begin{eqnarray*}
N_1&=&\frac{i}{\sqrt{2}}\xi_1 h_{\alpha_1}+\frac{i}{\sqrt{2}}\xi_2 h_{\alpha_2}- 
e_{\alpha_1}-e_{\alpha_2}-[e_{\alpha_2},e_{\alpha_0}] - [e_{\alpha_0}, e_{\alpha_1}]+
\frac{i}{\sqrt{2}}(\xi_1-\xi_2)e_{\alpha_0}\\ 
N_0&=&-\phi_1'h_{\alpha_1} - \phi_2'h_{\alpha_2}-(\phi_1'-\phi_2')e_{\alpha_0}.
\end{eqnarray*}
Hence,
\begin{eqnarray}
&&\mathcal{L}_B=\partial_u + (\frac{i}{\sqrt{2}}\xi_1h_{\alpha_1}+
\frac{i}{\sqrt{2}}\xi_2h_{\alpha_2}-[e_{\alpha_2},e_{\alpha_0}]-
[e_{\alpha_0},e_{\alpha_1}]\nonumber\\
&&-e_{\alpha_1}-e_{\alpha_2}+
\frac{i}{\sqrt{2}}(\xi_1-\xi_2)e_{\alpha_0})^2-(\phi'_1-\phi'_2)e_{\alpha_0}-
\phi'_1h_{\alpha_1}-\phi'_2h_{\alpha_2}.
\end{eqnarray}
Considering the associated linear problem, we can write the solution as
\begin{eqnarray}
\chi(u)=e^{\sum_{i=1,2}\phi_i(u)h_{\alpha_i}}Pexp\int_0^{u}\d u'(\sum_{k=0,1,2}
W_{\alpha_k}(u')e_{\alpha_k}+K(u'))
\end{eqnarray}  
where 
\begin{eqnarray}\label{K}
W_{\alpha_j}&=&\int \d\theta e^{-\Phi_j}, \quad j=1,2, \qquad  
W_{\alpha_0}=\int \d\theta (D\Phi_1-D\Phi_2)e^{\Phi_1+\Phi_2} \nonumber\\
&&K(u)=-\frac{i}{\sqrt{2}}\xi_2[e_{\alpha_2},e_{\alpha_0}]e^{\phi_2}-
\frac{i}{\sqrt{2}}\xi_1[e_{\alpha_0},e_{\alpha_1}]e^{\phi_1}-[e_{\alpha_1},e_{\alpha_2}]e^{-\phi_1-\phi_2}
-\nonumber\\
&&[[e_{\alpha_0},e_{\alpha_1}],[e_{\alpha_2},e_{\alpha_0}]]e^{\phi_1+\phi_2}-
[e_{\alpha_2},[e_{\alpha_0},e_{\alpha_1}]]-
[e_{\alpha_1},[e_{\alpha_2},e_{\alpha_0}]].
\end{eqnarray}
We can then define the monodromy matrix on the interval $[0,2\pi]$ with the quasiperiodic boundary conditions
$$\phi_i(u+2\pi)=\phi_i(u)+2\pi ip_i, \qquad \xi_i(u+2\pi)=\pm\xi_i(u)$$ 
as
\begin{eqnarray}\label{moncl}
\mathbf{M}=e^{\sum_{i=1,2}2i\pi p_ih_{\alpha_i}}Pexp\int_0^{2\pi}\d u(\sum_{k=0,1,2}
W_{\alpha_k}(u)e_{\alpha_k}+K(u)).
\end{eqnarray}  
The reason for separating the integrand into the $K$ term and the covariant part expressed in terms of
superfields is that 
in the quantum case (see below), the  
noncovariant $K$ term disappears from the expression for the quantum monodromy matrix.

As in the case of the standard KdV models, we define the auxiliary L-operators
\begin{eqnarray}\label{L}
\mathbf{L}=e^{-\sum_{i=1,2}i\pi p_ih_{\alpha_i}}\mathbf{M},
\end{eqnarray}  
which satisfy the quadratic r-matrix relation \cite{solitons}:  
\begin{eqnarray}
\{\mathbf{L}(\lambda),\otimes \mathbf{L}(\mu)\}=[\mathbf{r}(\lambda\mu^{-1}), 
\mathbf{L}(\lambda)\otimes\mathbf{L}(\mu)],
\end{eqnarray}
where we restore the dependence on the spectral parameters 
$\lambda$ and $\mu$ corresponding to some fixed evaluation representations. 
Here, $\mathbf{r}(\lambda\mu^{-1})$ 
is the classical trigonometric r-matrix associated with $sl^{(1)}(2|1)$ 
\cite{shadr}.
As usual, this yields a relation leading to the classical integrability,
\begin{eqnarray}
\{\mathbf{t}(\lambda),\mathbf{t}(\mu)\}=0,
\end{eqnarray}     
where $\mathbf{t}(\lambda)={\rm str} \mathbf{M}(\lambda)$ and the supertrace is 
taken in one of the $sl^{(1)}(2|1)$  representations.

\section{Quantum R-matrix and the Cartan-Weyl basis for $sl_q^{(1)}(2|1)$} 

The quantum algebra $sl_q^{(1)}(2|1)$ has the commutation relations
\begin{eqnarray}
&&[e_{\alpha_i}, e_{-\alpha_j}]=\delta_{i,j}[h_{\alpha_i}]\quad (i=0,1,2), \quad
[h_{\alpha_j},e_{\pm\alpha_i}]=\pm e_{\pm\alpha_i} \quad (i=1,2, i \neq j)\nonumber\\
&&[h_{\alpha_j},e_{\pm\alpha_0}]=\mp e_{\pm\alpha_0} \quad (i=1,2),\quad
[h_{\alpha_0},e_{\pm\alpha_i}]=\mp e_{\pm\alpha_i} \quad (i=1,2)\nonumber\\
&&[h_{\alpha_i},e_{\pm\alpha_i}]=0 \quad (i=1,2),\quad
[h_{\alpha_0},e_{\pm\alpha_0}]=\pm 2e_{\pm\alpha_0},\nonumber\\
&&\label{serre}[e_{\pm\alpha_i},[e_{\pm\alpha_i},e_{\pm\alpha_j}]_q]_q=0,\quad e^2_{\alpha_k}=0 \quad (k=1,2).
\end{eqnarray} 
where $[x]=(q^x-q^{-x})/(q-q^{-1})$ and 
the quantum supercommutator is defined as
$$[e_{\alpha},e_{\beta}]_q=
e_{\alpha}e_{\beta}-(-1)^{p(\alpha)p(\beta)}q^{(\alpha,\beta)}e_{\beta}e_{\alpha}.
$$ 
The universal quantum R-matrix for quantum affine superalgebras is 
\cite{khor}
\begin{equation}\label{R-matrix}
R=K\bar{R}=K(\prod^{\to}_{\alpha\in{\Delta_{+}}}R_{\alpha}),
\end{equation}
where $\bar{R}$ is a reduced R-matrix and $R_{\alpha}$ are defined by the formulas
\begin{equation}
R_{\alpha}=exp_{q_{\alpha}^{-1}}((-1)^{p(\alpha)}(q-q^{-1})(a(\alpha))^{-1}(e_{\alpha}\otimes e_{-\alpha}))
\end{equation}
for real roots and 
\begin{equation}
R_{n\delta}=exp((q-q^{-1})(\sum^{mult}_{i,j}c_{ij}(n)
e^{(i)}_{n\delta}\otimes e^{(j)}_{-n\delta}))
\end{equation}
for pure imaginary roots. 
Here, $\Delta_{+}$ is the reduced positive root system (the bosonic roots  equal to twice 
a fermionic root are excluded) and q-exponen\-tial is defined as usual, 
$$ 
exp_q(x)=
\sum^{\infty}_{n=0} x^n/(n)_q!, \qquad
(n)_q=(q^n-1)/(q-1).
$$ 
The generators corresponding to composite roots and the ordering in 
(\ref{R-matrix}) are defined in accordance with the construction of the Cartan-Weyl basis 
(see, e.g., \cite{khor}). 
The Cartan factor $K$ in the case of the $sl_q^{(1)}(1|2)$ is equal to 
$
K=q^{h_{\alpha_1}\otimes h_{\alpha_2}+h_{\alpha_2}\otimes h_{\alpha_1}}
$. 
The coefficients $a(\alpha), c_{ij}(n)$ and $d_{ij}(n)$ are defined using relations
$$
[e_{\gamma},e_{-\gamma}]=a(\gamma)(k_{\gamma}-k^{-1}_{\gamma})/(q-q^{-1}), 
$$ 
$$[e^{(i)}_{n\delta},e^{(j)}_{-n\delta}]=d_{ij}(n)(q^{nh_{\delta}}-q^{-nh_{\delta}})/(q-q^{-1})$$ 
and 
$c_{ij}(n)$ is the matrix inverse to $d_{ij}(n)$.
The first few generators corresponding to composite roots are constructed using the 
above procedure as
\begin{eqnarray}
&&e_{\alpha_1+\alpha_2}=[e_{\alpha_1},e_{\alpha_2}]_{q^{-1}}\\
&&e_{\delta-\alpha_1}\equiv e_{\alpha_0+\alpha_2}= [e_{\alpha_0},e_ {\alpha_2}]_{q^{-1}},\quad 
e_{\delta-\alpha_2}\equiv e_{\alpha_0+\alpha_1}= [e_{\alpha_1},e_ {\alpha_0}]_{q^{-1}}\nonumber\\
&&e^{(1)}_{\delta}=[[e_{\alpha_0},e_ {\alpha_2}]_{q^{-1}},e_{\alpha_1}],\quad 
e^{(2)}_{\delta}=[[e_{\alpha_1},e_ {\alpha_0}]_{q^{-1}},e_ {\alpha_2}]\nonumber\\
&&e_{2\delta-\alpha_1-\alpha_2}=[e_{\delta-\alpha_2},e_{\delta-\alpha_1}]_{q^{-1}}\nonumber\\
&&e_{-\alpha_1-\alpha_2}=[e_{-\alpha_2},e_{-\alpha_1}]_{q}\nonumber\\
&&e_{-\delta+\alpha_1}\equiv e_{-\alpha_0-\alpha_2}= [e_{-\alpha_2},e_ {-\alpha_0}]_{q},\quad 
e_{-\delta+\alpha_2}\equiv e_{-\alpha_0-\alpha_1}= [e_{-\alpha_0},e_ {-\alpha_1}]_{q}\nonumber\\
&&e^{(1)}_{-\delta}=[[e_{-\alpha_2},e_ {-\alpha_0}]_{q},e_{-\alpha_1}],\quad 
e^{(2)}_{-\delta}=[[e_{-\alpha_0},e_ {-\alpha_1}]_{q},e_{-\alpha_2}]\nonumber\\
&&e_{-2\delta+\alpha_1+\alpha_2}=[e_{-\delta+\alpha_1},e_{-\delta+\alpha_2}]_{q}.\nonumber
\end{eqnarray}

%%%%%%%%%%%%%%%%%%%%%%%%%%%%%%%%%%%%%%%%%%%%%%%%%%%%%%
\section{Construction of the Quantum Monodromy Matrix}

In this section, we construct the quantum version of the monodromy matrix introduced in Sec. 2. 
We show that matrix (\ref{moncl})  is reproduced in the classical limit.

We first consider the quantum versions of the free-field exponentials (vertex operators):
\begin{eqnarray}\label{vertex}
&&W_{\alpha_i}=\int \d\theta :e^{-\Phi_i}:\equiv \frac{i}{\sqrt{2}}\xi_i :e^{-\phi_i}: \quad(i=1,2),\nonumber\\
&&W_{\alpha_0}=\int \d\theta :(D\Phi_1-D\Phi_2)e^{\Phi_1+\Phi_2}:\equiv :e^{\phi_1+\phi_2}(\phi'_1-
\phi'_2+\xi_1\xi_2):.
\end{eqnarray}
We can express the superfields as $\Phi_1=\frac{i\Phi_{+}+\Phi_{-}}{\sqrt{2}}$ and
$\Phi_2=\frac{i\Phi_{+}-\Phi_{-}}
{\sqrt{2}}$, where 
\begin{eqnarray}
&&\phi_{\pm}(u)=iQ^{\pm}+iP^{\pm}u+\sum_n\frac{a^{\pm}_{-n}}{n}e^{inu},\qquad
\xi_{\pm}(u)=i^{-1/2}\sum_n\xi^{\pm}_ne^{-inu},\nonumber\\
&&[Q^{\pm},P^{\pm}]=\frac{i}{2}\beta^2 ,\quad 
[a^{\pm}_n,a^{\pm}_m]=\frac{\beta^2}{2}n\delta_{n+m,0},\quad
\{\xi^{\pm}_n,\xi^{\pm}_m\}=\beta^2\delta_{n+m,0}.
\end{eqnarray}
and the normal ordering in (\ref{vertex}) is defined as 
$$
:e^{c\phi_{\pm}(u)}:=
\exp\Big(c\sum_{n=1}^{\infty}\frac{a^{\pm}_{-n}}{n}e^{inu}\Big)
\exp\Big(ci(Q^{\pm}+P^{\pm}u)\Big)\exp\Big(-c\sum_{n=1}^{\infty}\frac{a^{\pm}_{n}}{n}e^{-inu}.
\Big).
$$
Here, the $a^{\pm}_{n}$ operators with negative $n$ are placed to the left and those with positive $n$,
 to the right.

Vertex operators (\ref{vertex}) integrated from $u_1$ to $u_2$ satisfy the quantum Serre and 
``nonstandard'' Serre relations  for the lower Borel subalgebra with 
$q=e^{{i\pi\beta^2}/{2}}$. 
Proving  this is nontrivial because the usual proof of the Serre relations 
given in  \cite{feigin} for the bosonic case, based on  transforming 
the product of the integrated vertex operators to  ordered integrals, is inapplicable here 
because of the singularities generated by the fermion fields in the corresponding 
operator product expansions. But there is another way to prove it, relying on the standard
conformal field theory technique of 
contour integration and analytic continuation of operator product expansions of nonlocal vertex 
operators \cite{klevtsov}. 
This proof is also applicable to a quantum affine superalgebra and the corresponding 
vertex operators because this method allows isolating the divergences  in each product of 
vertex operators 
and then canceling 
them in the standard  and ``nonstandard'' Serre relations. This will be considered  
elsewhere for a general quantum affine superalgebra.     

Because the operators $(q-q^{-1})^{-1}\int^{u_2}_{u_1}W_{\alpha_i}$ satisfy the Serre relations, 
we can represent the lower Borel subalgebra using the correspondence
$$
e_{-\alpha_i}  \to (q-q^{-1})^{-1}\int^{u_2}_{u_1}W_{\alpha_i}.
$$
 It was shown in \cite{bhk} that the 
corresponding reduced R-matrix $\bar{R}$, denoted by 
$\bar{\mathbf{L}}^{(q)}(u_2,u_1)$ here, has  
the P-exponential property, satisfying the functional relation  
\begin{eqnarray}
\bar{\mathbf{L}}^{(q)}(u_3,u_1)=\bar{\mathbf{L}}^{(q)}(u_3,u_2)
\bar{\mathbf{L}}^{(q)}(u_2,u_1).
\end{eqnarray}
But in the supersymmetric case involving  fermionic operators, the associated 
singularities in the operator products do not allow  writing $\bar{\mathbf{L}}^{(q)}(u_3,u_1)$ 
in the standard manner in terms of ordered integrals. 
We therefore call it  the ``quantum'' P-exponential. In our case it can be written as
\begin{eqnarray}
\bar{\mathbf{L}}^{(q)}(u_2,u_1)=Pexp^{(q)}\int_{u_1}^{u_2}\d u(\sum_{k=0,1,2}
W_{\alpha_k}(u)e_{\alpha_k}).
\end{eqnarray}
It can be shown that the operators $e^{\sum_{i=1,2}\pi i p_ih_{\alpha_i}}\bar{\mathbf{L}}^{(q)}(2\pi,0)=
\mathbf{L}^{(q)}$ satisfy the RTT-relation \cite{leshouches}:
\begin{eqnarray}
&&\mathbf{R}(\lambda\mu^{-1})
\Big(\mathbf{L}^{(q)}(\lambda)\otimes \mathbf{I}\Big)\Big(\mathbf{I}
\otimes \mathbf{L}^{(q)}(\mu)\Big)\\
&&=(\mathbf{I}\otimes \mathbf{L}^{(q)}(\mu)\Big)
\Big(\mathbf{L}^{(q)}(\lambda)\otimes \mathbf{I}\Big)\mathbf{R}
(\lambda\mu^{-1}), \nonumber
\end{eqnarray}
where the dependence on $\lambda$ and $\mu$ means that we consider
$\mathbf{L}^{(q)}$-operators in the appropriate evaluation representation of $sl_q^{(1)}(2|1)$.
We also note  that 
$$\mathbf{M}^{(q)}=e^{\sum_{i=1,2}\pi i p_ih_{\alpha_i}}\bar{\mathbf{L}}^{(q)}
$$ 
satisfies the reflection equation \cite{reflection}
\begin{eqnarray}
\mathbf{\tilde{R}}_{12}(\lambda\mu^{-1})
\mathbf{M}_1^{(q)}(\lambda)F^{-1}_{12}\mathbf{M}^{(q)}_2(\mu)=\mathbf{M}_2^{(q)}(\mu)F^{-1}_{12}
\mathbf{M}_1^{(q)}(\lambda)\mathbf{R}_{12}(\lambda\mu^{-1}),
\end{eqnarray}
where $F=K^{-1}$, the inverse  Cartan factor from the universal R-matrix, 
$$
\mathbf{\tilde{R}}_{12}(\lambda\mu^{-1})=F^{-1}_{12}\mathbf{R}_{12}(\lambda\mu^{-1})F_{12}
$$ 
and labels 1 and 2 indicate the position of factors in the tensor 
product. This leads to the quantum integrability relation
\begin{eqnarray}
[\mathbf{t}^{(q)}(\lambda), \mathbf{t}^{(q)}(\mu)]=0,
\end{eqnarray} 
where $\mathbf{t}^{(q)}(\lambda)={\rm str} \mathbf{M}^{(q)}(\lambda)$.

We next show that $\mathbf{L}^{(q)}$ and  $\mathbf{M}^{(q)}$ pass into
the respective auxiliary $\mathbf{L}$-matrix and  monodromy matrix 
in the classical limit as $q\to 1$.
We also note that the quantum universal R-matrix, as usual, tends to 
the classical r-matrix in the limit as $q\to 1$, and the classical limit of the 
$\mathbf{L}^{(q)}$-operator therefore gives a realization 
of the classical r-matrix via the classical counterparts of the corresponding vertex operators.

To find the classical limit, 
we decompose $\mathbf{\bar{L}}^{(q)}$ as \cite{toda-kdv}:
\begin{equation}
\mathbf{\bar{L}}^{(q)}(2\pi,0)=
\lim_{N\to\infty}\prod_{m=1}^{N}\mathbf{\bar{L}}^{(q)}(x_{m},x_{m-1}), 
\end{equation}
where we divide the interval $[0,2\pi]$ into infinitesimal 
intervals $[x_m,x_{m+1}]$
with $x_{m+1}-x_m=\epsilon=2\pi/N$.
We nexi find the terms that can  contribute to
$\mathbf{\bar{L}}^{(q)}(x_{m},x_{m-1})$ in the first order 
in $\epsilon$. 
For this, we need the operator product expansions of 
vertex operators. Nontrivial terms occur in the expansions
\begin{eqnarray}
&&\xi_1(u)\xi_2(u')=
\frac{i\beta^2}{(iu-iu')}+\sum_{k=0}^{\infty}c_k(u)(iu-iu')^k,\nonumber\\
&&:e^{a\phi_1(u)}::e^{b\phi_2(u')}:=(iu-iu')^{\frac{-ab\beta^2}{2}}
(:e^{(a\phi_1(u)+b\phi_2(u)}:+\nonumber\\
&&\sum_{k=1}^{\infty}d_k(u)(iu-iu')^k),\nonumber\\
&&\phi_1'(u):e^{b\phi_2(u')}:=\frac{-ib\beta^2:e^{b\phi_2(u)}:}{2(iu-iu')}+
\sum_{k=0}^{\infty}f_k(u)(iu-iu')^k,\nonumber\\
&&\phi_2'(u):e^{b\phi_1(u')}:=\frac{-ib\beta^2:e^{b\phi_1(u)}:}{2(iu-iu')}+
\sum_{k=0}^{\infty}f_k(u)(iu-iu')^k.
\end{eqnarray}
In the cases considered in \cite{toda-kdv}, only two types of terms 
contribute to $\mathbf{\bar{L}}^{(q)}(x_{m-1},x_{m})$ in the order $\epsilon$ as  $q\to 1$.
Terms of
the first type are operators of the first order in $W_{\alpha_i}$, 
and terms of the second type are
the operators quadratic in $W_{\alpha_i}$, which contribute in the order
$\epsilon^{1\pm\beta^2}$ by virtue of operator product expansion. These second-type contributions correspond
 to the composite roots that are equal to the sum of two simple roots.

In this paper, we show that there are contributions of the composite roots  
equal to the sum of three and even four simple roots, ensuring the desired terms in the classical 
expression (\ref{K}), (\ref{moncl}).
We first consider the quadratic terms corresponding to the negative roots $-\alpha_1-\alpha_2$, 
$-\delta+\alpha_2$ and $-\delta+\alpha_1$. The commutation relations between vertex operators on a circle are
such that for $u>u'$,
\begin{eqnarray}
&&W_{\alpha_i}(u)W_{\alpha_j}(u')=-q^{-1}W_{\alpha_j}(u')W_{\alpha_i}(u)
\quad u>u' \quad (i,j=1,2\quad i\neq j)\nonumber\\
&&W_{\alpha_i}(u)W_{\alpha_0}(u')=q W_{\alpha_0}(u')W_{\alpha_i}(u) 
\quad u>u' \quad (i=1,2) \nonumber\\
&&W_{\alpha_0}(u)W_{\alpha_i}(u')=q W_{\alpha_i}(u')W_{\alpha_0}(u), 
 \quad i=1,2. 
\end{eqnarray}
This allows writing the generators corresponding to the negative composite roots  
$-\delta+\alpha_2$ and $-\delta+\alpha_1$ as
\begin{eqnarray}
&&[e_{-\alpha_0}, e_{-\alpha_1}]_q=
\frac{1}{q-q^{-1}}\int_{x_{m-1}}^{x_{m}}\d uW_{\alpha_1}(u)\int_{x_{m-1}}^u 
\d u'W_{\alpha_0}(u'),\nonumber\\
&&[e_{-\alpha_2}, e_{-\alpha_0}]_q=
\frac{1}{q-q^{-1}}\int_{x_{m-1}}^{x_{m}}\d uW_{\alpha_0}(u)\int_{x_{m-1}}^u \d u'W_{\alpha_2}(u').
\end{eqnarray}
The exponents of the corresponding q-exponentials in quantum R-matrix (\ref{R-matrix}) are equal to  
\begin{eqnarray}
&&\int_{x_{m-1}}^{x_{m}}\d uW_{\alpha_1}(u)\int_{x_{m-1}}^u \d u'W_{\alpha_0}(u')[e_{\alpha_1}, e_{\alpha_0}]_{q^{-1}},\nonumber\\
&&\int_{x_{m-1}}^{x_{m}}\d uW_{\alpha_0}(u)\int_{x_{m-1}}^u \d u'W_{\alpha_2}(u')[e_{\alpha_0}, e_{\alpha_2}]_{q^{-1}}.
\end{eqnarray}
In the classical limit ($\beta^2\to 0$), their contribution calculated using 
the corresponding operator product expansions is given by 
\begin{eqnarray}
\int_{x_{m-1}}^{x_{m}}\d u(-\frac{i}{\sqrt{2}}\xi_2e^{\phi_2}[e_{\alpha_2},e_{\alpha_0}]-
\frac{i}{\sqrt{2}}\xi_1e^{\phi_1}[e_{\alpha_0},e_{\alpha_1}])
\end{eqnarray}
(similarly to \cite{super-kdv}). 
The contribution of the root $-\alpha_1-\alpha_2$ is 
\begin{eqnarray}
\int_{x_{m-1}}^{x_{m}}\d u(-e^{-\phi_1-\phi_2}[e_{\alpha_1},e_{\alpha_2}]).
\end{eqnarray}
In this case, the commutator of the integrated vertex operators cannot be rewritten in terms of  ordered 
integrals as above. We use the fact that the integrated vertex operators must be radially ordered, 
i.e., the product $e_{-\alpha_i}e_{-\alpha_j}$, for example, must be written as 
$$
\int^{x_{m}}_{x_{m-1}}\d uW_{\alpha_i}(u-i0)
\int^{x_{m}}_{x_{m-1}}\d u' W_{\alpha_i}(u'+i0),
$$
 and recall the well known relation 
\begin{equation}
\frac{1}{x+i0}-\frac{1}{x-i0}=-2i\pi\delta(x).
\end{equation}
Calculations performed as a generalization of the results in  \cite{toda-kdv} lead to (35).
At the ``quadratic'' level, we therefore have complete agreement with the classical 
expression.

We now consider  the contribution corresponding to the composite roots $e^{(i)}_{\delta}$ and 
$2\delta-\alpha_1-\alpha_2$. 
We first  consider the purely imaginary roots, 
$$
e^{(1)}_{-\delta}=[[e_{-\alpha_2},e_ {-\alpha_0}]_{q},e_{-\alpha_1}], \qquad 
e^{(2)}_{-\delta}=[[e_{-\alpha_0},e_ {-\alpha_1}]_{q},e_ {-\alpha_2}].
$$
Finding their contribution in the 
classical limit requires calculating the  contribution to
 $[e_{\alpha_2},e_ {\alpha_0}]_{q}$ and $[e_{\alpha_0},e_ {\alpha_1}]_{q}$ 
of the terms of the order  $\epsilon^{1+\beta^2}$, 
proportional to $\int \d u \d\theta e^{\Phi_2}$ and  $\int \d u \d\theta e^{\Phi_1}$. 
We then consider their supercommutators with  $e_{\alpha_1}$ and $e_ {\alpha_2}$. 
Rewriting these supercommutators in terms of ordered integrals as above and following the 
calculations in \cite{super-kdv}, we find that the contribution of the associated exponent of q-exponential 
is given by
\begin{eqnarray}
\int^{x_{m}}_{x_{m-1}}\d u(-[e_{\alpha_2},[e_{\alpha_0},e_{\alpha_1}]]-
[e_{\alpha_1},[e_{\alpha_2},e_{\alpha_0}]]).
\end{eqnarray}
The generator
 $e_{2\delta-\alpha_1-\alpha_2}$  is expressed similarly to the q-commutator of 
$[e_{\alpha_2},e_ {\alpha_0}]_{q}$  and $[e_{\alpha_0},e_ {\alpha_1}]_{q}$.
Taking the terms of the order $\epsilon^{1+\beta^2}$ into account and 
using the formula (36) as explained above, we obtain the order-$\epsilon$
contribution to the classical expression
\begin{eqnarray}
\int^{x_m}_{x_{m-1}}\d u(-[[e_{\alpha_0},e_{\alpha_1}],[e_{\alpha_2},e_{\alpha_0}]]e^{\phi_1+\phi_2}).
\end{eqnarray}
Gathering all the terms, we thus find  that the first iteration of the $\mathbf{\bar{L}}$-operator
in the classical limit is given by
\begin{eqnarray}
&&\lim_{q\to 1}\mathbf{\bar{L}}^{(q)}(x_{m},x_{m-1})=1+\int^{x_m}_{x_{m-1}}\d u 
(\sum_{k=0,1,2}W_{\alpha_k}(u)e_{\alpha_k}-\frac{i}{\sqrt{2}}\xi_2[e_{\alpha_2},e_{\alpha_0}]e^{\phi_2}\nonumber\\
&&-
\frac{i}{\sqrt{2}}\xi_1[e_{\alpha_0},e_{\alpha_1}]e^{\phi_1}-[e_{\alpha_1},e_{\alpha_2}]e^{-\phi_1-\phi_2}
-[[e_{\alpha_0},e_{\alpha_1}],[e_{\alpha_2},e_{\alpha_0}]]e^{\phi_1+\phi_2}-\nonumber\\
&&[e_{\alpha_2},[e_{\alpha_0},e_{\alpha_1}]]-
[e_{\alpha_1},[e_{\alpha_2},e_{\alpha_0}]])
+O(\epsilon^2).
\end{eqnarray}
Collecting all the infinitesimal $\bar{\mathbf{L}}$ operators and multiplying them by the appropriate Cartan factor, 
we obtain the sought classical expressions for the $\mathbf{L}$-operators and the monodromy matrix. 

\section{Conclusions}

The construction of the monodromy matrix given in the previous section is based on  the method
proposed  in 
\cite{super-kdv}-\cite{blz}, \cite{bhk}. 
In other words, we have shown that
 the quantum version of the auxiliary $\mathbf{L}$-matrix  
coincides with the universal R-matrix with the lower Borel algebra represented by the appropriate vertex operators.
This construction also allows showing that the supersymmetry generators commute with the supertrace of the 
monodromy matrix and can therefore be included in the series of integrals of motion. 
Indeed, the commutators of the supersymmetry generators
\begin{eqnarray}
G^+_0=\beta^{-2}\sqrt{2}i^{-1/2}\int_0^{2\pi}du \phi_1'(u) \xi_2(u), \quad 
G^-_0=\beta^{-2}\sqrt{2}i^{-1/2}\int_0^{2\pi}du \phi_2'(u) \xi_1(u)\nonumber
\end{eqnarray}
with vertex operators reduce to total derivatives; the argument then repeads Sec. 4 in 
\cite{toda-kdv2}, where it was shown (following \cite{bhk})
that in the case of the standard Drinfeld-Sokolov hierarchies, 
the supersymmetry generator commutes with the trace of the monodromy 
matrix if the system of  simple roots is purely fermionic (the oddness condition 
for the system of simple roots then
guarantees that the commutators of supersymmetry generators 
with the corresponding vertex operators are equal to  total derivatives).
It can be shown similarly 
that the transfer matrices commute with the zero mode of the 
U(1) current of 
 the $N=2$ superconformal algebra (the quantum version of the $V$ field in (\ref{poisson})).
  
This result has  important consequences. If we make the twist 
transformation \cite{twist} in the 
underlying N=2 superconformal algebra, then we find that one of the generators $G^{\pm}_0$ becomes 
a BRST operator. 
That is, the transfer-matrices become BRST exact, providing an infinite 
series of pairwise 
commuting ``physical'' integrals of 
motion (of zero ghost number, as follows from the commutation relations with
the zero mode of U(1) current). 
This allows  studying  two-dimensional topological models and their 
integrable perturbations using
the methods of integrable theories, for example,
the well-known quantum inverse scattering method 
\cite{leshouches}. 

\section*{Acknowledgements}
The author is grateful to the referee for the useful comments and is especially 
grateful to N. Reshetikhin, M. Semenov-Tian-Shansky, and 
A. Tseytlin for encouragement.

This work was supported  by the Dynasty foundation, 
the CRDF (Grant No. RUM1-2622-ST-04), and
the Russian Foundation for Basic Research (Grant No. 05-01-00922).

\end{document}